\documentclass[preprint,sd,amsmath,amssymb,reprint,graphicx]{revtex4-1}

\usepackage{graphicx}
\usepackage{dcolumn}
\usepackage{bm}
\usepackage{SIunits}

\draft 

\begin{document}


\title{Annealing shallow Si/SiO$_2$ interface traps in electron-beam irradiated high-mobility metal-oxide-silicon transistors} 



\author{J.-S. Kim}
\email[]{jk3@princeton.edu}
\author{A. M. Tyryshkin}
\email[]{atyryshk@princeton.edu}
\author{S. A. Lyon}
\email[]{lyon@princeton.edu}
\affiliation{Department of Electrical Engineering, Princeton University, Princeton, NJ 08544, USA}


\date{\today}

\begin{abstract}
Electron-beam (e-beam) lithography is commonly used in fabricating metal-oxide-silicon (MOS) quantum devices but creates defects at the Si/SiO$_2$ interface. Here we show that a forming gas anneal is effective at removing shallow defects ($\leq4$ meV below the conduction band edge) created by an e-beam exposure by measuring the density of shallow electron traps in two sets of high-mobility MOS field-effect transistors (MOSFETs). One set was irradiated with an electron-beam (10 keV, 40 $\micro$C/cm$^2$) and was subsequently annealed in forming gas while the other set remained unexposed. Low temperature (335 mK) transport measurements indicate that the forming gas anneal recovers the e-beam exposed sample's peak mobility (14,000 cm$^2$/Vs) to within a factor of two of the unexposed sample's mobility (23,000 cm$^2$/Vs). Using electron spin resonance (ESR) to measure the density of shallow traps, we find that the two sets of devices are nearly identical, indicating the forming gas anneal is sufficient to anneal out shallow defects generated by the e-beam exposure. Fitting the two sets of devices' transport data to a percolation transition model, we extract a T=0 percolation threshold density in quantitative agreement with our lowest temperature ESR-measured trap densities.

\end{abstract}


\maketitle 


Recent work on metal-oxide-silicon (MOS) quantum devices has demonstrated superb control of single electrons in electrostatically defined quantum dots.\cite{morello,dzurak,nordbergapl,jiang2014} Silicon structures are a promising platform for the realization of a quantum processor,\cite{loss,kane} demonstrating long spin coherence times\cite{tyryshkin,morello,zwanenburg} and a large valley splitting.\cite{takashina,yang,zwanenburg} In addition, silicon enjoys a mature fabrication infrastructure thanks to the complementary MOS industry and is compatible with single ion implantation,\cite{morello2012,singhsb,harveycollard} allowing for electron-donor coupled qubits.\cite{harveycollard, pica} One of the most pressing difficulties in fabricating MOS quantum devices is the presence of disorder at the Si/SiO$_2$ interface,\cite{zwanenburg} specifically shallow defects a few meV below the conduction band edge ($E_C$). These shallow defects are of the same energy scale of a typical electrostatically defined quantum dot potential, and are electrically active at the operating temperatures of a quantum dot device ($<1$ K).\cite{nordbergapl,morello,jiang,dzurak,zwanenburg} In contrast, electrons confined in deeper, mid-gap states are frozen in place and are inert, contributing a static electric field background. As a result, the presence of shallow defects can be catastrophic for single electron control in quantum devices operating at low temperature. These interface traps can inadvertently be introduced during device fabrication by high energy processes, especially from e-beam lithography,\cite{nordberg} the ``workhorse" of quantum device fabrication in research labs.\cite{morello,dzurak,nordbergapl,jiang2014,harveycollard,jiang,nordberg} While a large body of literature exists on annealing irradiated Si/SiO$_2$ interface defects,\cite{nicbrews,tpma} most\cite{frystak} of these measurements are done well above liquid helium temperatures and characterize defects far away from the conduction band edge. Thus, little is known about annealing shallow traps which only manifest themselves at low temperatures. 

We have fabricated devices with the highest reported electron mobility for a MOSFET with an oxide thickness of 30 nm or thinner\cite{ando,zwanenburg} (23,000 cm$^2$/Vs) and subjected these devices to e-beam irradiation and a subsequent forming gas anneal. Using electron spin resonance (ESR),\cite{jock,shankar} we directly measure the shallow trap density of our devices and compare these results to more typical measurements of the Si/SiO$_2$ interface, namely, low temperature transport measurements of electron mobility\cite{ando,zwanenburg,nordberg} and percolation thresholds.\cite{tracy} Measurements of electron mobility are a commonly used method of assessing the oxide interface in MOS devices but provide only indirect measurements of confined shallow traps.\cite{jock} Our transport data show that the e-beam dose significantly degrades a device's peak mobility, but a forming gas anneal can restore its peak mobility to within a factor of two of the unexposed sample's. Despite this difference in peak mobility, our devices display very similar T=0 percolation threshold densities. Our ESR measurements of the density of shallow traps demonstrate that a forming gas anneal effectively removes shallow traps generated in the e-beam exposed sample over the entire measured energy range of approximately 4 meV (4.2 K) to 0.3 meV (360 mK) below the conduction band edge. We find our devices' lowest temperature ESR measurement of shallow traps match the T=0 percolation threshold density, demonstrating agreement between two independent methods of assessing the oxide interface.

The devices measured in this work are n-channel inversion MOSFETs fabricated at Princeton from a commercially grown (Novati Technologies) gate stack. Our starting substrate consists of a high resistivity (1000-3000 $\ohm$-cm) float zone p-type (100) silicon wafer with 30 nm of dry, chlorinated thermal oxide and capped with 200 nm of un-doped amorphous silicon (a-Si). For both sets of devices, large-area MOSFETs ($3.3\times20$ mm$^2$) were fabricated for ESR measurements and Hall bars ($0.2\times4$ mm$^2$) were fabricated for transport measurements. A large gate area ($\sim1$ cm$^2$) is necessary to detect the spin signal of 2D electrons at 4.2 K using X-band ESR.\cite{shankar} Holes were etched through the a-Si for self aligned ohmic contacts using an SF$_6$ and C$_4$F$_8$ based plasma and the underlying oxide was etched with buffered HF. The devices were implanted with As (35 keV, 5 $\times$ 10$^{15}$ cm$^{-2}$, Leonard Kroko, Inc.) to dope the source/drain contacts as well as the a-Si gate. The n$^+$ a-Si was then etched in SF$_6$ and C$_4$F$_8$ to define the gate geometry. The samples were cleaned in an O$_2$ plasma for 10 minutes, RCA cleaned in quartz dishes, and annealed in N$_2$ (900 $\degree$C, 1 hour) to activate the implanted dopants, crystallize the n$^+$ a-Si into poly-silicon, and to reduce fixed charge at the Si/SiO$_2$ interface.\cite{nicbrews}

\begin{figure}
\includegraphics[width=\linewidth]{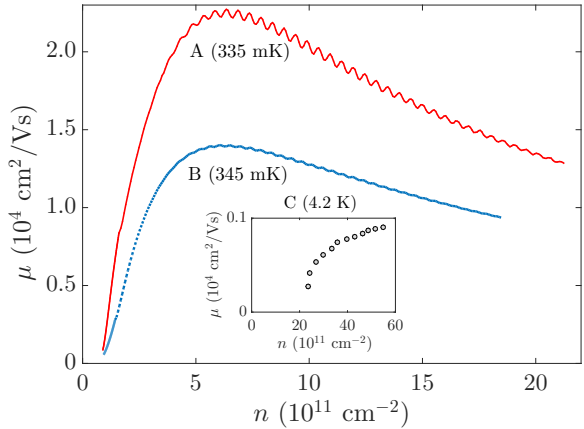}
\caption{\label{mobility}Experimental mobility ($\mu$) plotted as a function of electron density ($n$) for three studied devices. Sample A remained unexposed to the e-beam. Sample B was exposed and annealed in forming gas, recovering more than half of sample A's peak mobility. Sample C (inset) was exposed with no post-exposure anneal and shows significant degradation to its peak $\mu$ and a high threshold $n$. Shubnikov-de Haas oscillations are visible in samples A and B.}
\end{figure}

At this point the process was split and one set of devices (samples B) received a blanket e-beam exposure (10 keV, 40 $\micro$C/cm$^2$) and the other set (samples A) remained unexposed. The e-beamed samples were spin coated with 300 nm of ZEP520a e-beam resist prior to irradiation to ensure proper alignment of the exposure. The calculated penetration depth of 10 keV electrons through our resist and device stack is $\approx 1$ $\micro$m (Casino\cite{casino}), deep enough to damage the Si/SiO$_2$ interface. After stripping the e-beam resist, both sets of devices were metallized with 300 nm of thermally evaporated Al over the source/drain and n$^+$ poly-Si gate. Al was evaporated over the gate to avoid loading the ESR resonator by the resistive poly-silicon\cite{lo} and to accelerate the passivation of interface defects.\cite{plummer} Thermal evaporations were chosen over e-beam evaporations in this process as the latter method creates X-rays that can further damage the Si/SiO$_2$ interface.\cite{nordberg} Both sets of devices were then annealed in forming gas (5$\%$ H$_2$, 435$\degree$C) for 25 minutes. Finally, Ti/Au was thermally evaporated onto the contacts for soldering to the device. In addition to the two sets of devices mentioned above, a third Hall bar (sample C) was fabricated to demonstrate the damage created by the e-beam exposure. This sample was fabricated identically to sample A but was then coated with e-beam resist and received an e-beam exposure (identical to sample B) at the very end of processing, with no post-exposure anneal. 

\begin{figure}
\includegraphics[width=\linewidth]{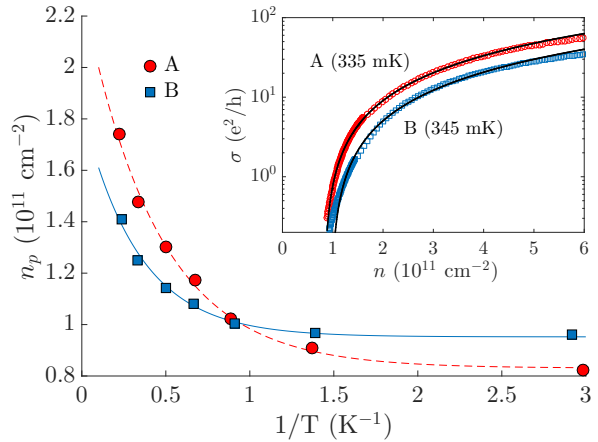}
\caption{\label{percolation} Percolation threshold density ($n_p$) for samples A and B as a function of temperature. Fits to the form $n_p=n_0+Ce^{-b/T}$ are also shown. Inset shows conductivity data ($\sigma$) vs. $n$ and percolation fits with $p=1.31$ to extract $n_p$ for sample A at 335 mK and sample B at 345 mK.}
\end{figure}

Transport measurements were done on all three sets of devices using standard low frequency lock-in techniques. Samples A and B were measured in a $^3$He cryostat (Janis Research) at temperatures between 335 mK and 4.5 K using a constant excitation current of 1.5 nA. The threshold voltage ($V_{th}$) of these two devices was measured to be $\approx 0.07$ V at 4.2 K and increases slightly with decreasing temperature to $\approx 0.2$ V at 335 mK. Sample C was measured at 4.2 K with an excitation current of 115 nA and its threshold voltage was measured to be 1.2 V, indicating approximately 10$^{12}$ cm$^{-2}$ net oxide charges (interface states and fixed oxide charge) created by the e-beam exposure. For each sample, electron densities ($n$) were calibrated by measuring the Hall resistivity in a 0.5 T field and the mobility was extracted by standard four-terminal lock-in measurements of the sample resistivity. Mobility ($\mu$) data are summarized in Figure \ref{mobility} which shows a peak mobility for sample A of 23,000 cm$^2$/Vs at $n=6.3\times10^{11}$ cm$^{-2}$, 14,000 cm$^2$/Vs at $n=6.1\times10^{11}$ cm$^{-2}$ for sample B, and $<1,000$ cm$^2$/Vs at $n=5.5\times10^{12}$ cm$^{-2}$ for sample C. Sample A demonstrates the highest reported electron mobility for a MOSFET with an oxide thickness of 30 nm or thinner.\cite{ando,zwanenburg} Comparing the mobility data for all three devices show that the e-beam dose significantly degrades the oxide interface (sample C) and that a forming gas anneal is sufficient to restore an e-beamed device to high mobility (sample B). 

\begin{figure}
\includegraphics[width=\linewidth]{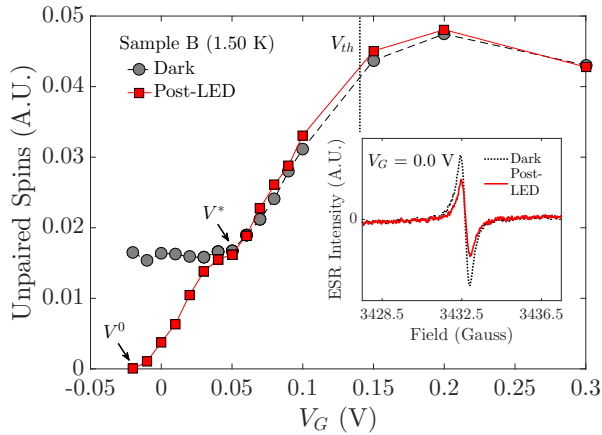}
\caption{\label{esr} Total number of unpaired spins plotted as a function of gate voltage for sample B at 1.50 K as measured in the dark (circles) and after above gap illumination (squares). $V_{th}=0.14$ V at this temperature, shown by the dashed vertical line. Inset: ESR spectra at $V_G=0.0$ V measured in the dark and after above gap illumination.}
\end{figure}

Peak mobility, however, is measured at relatively high electron densities where the two dimensional electron gas (2DEG) can effectively screen out scattering centers\cite{ando} and as such is not necessarily a useful indicator of the oxide interface quality for quantum devices operating in the few electron regime.\cite{jock} An alternative method used to assess the interface quality from transport measurements is to fit the measured conductivity ($\sigma$) to a percolation transition model\cite{tracy} of the form $\sigma(n)=A (n-n_p)^p$, and extract the percolation threshold density ($n_p$). $n_p$ gives a measure of the minimum number of carriers required to fill the disorder landscape before a conducting pathway can be supported. Holding the critical percolation exponent $p$ at 1.31, the expected value for a 2D system,\cite{tracy} and fixing the pre-factor $A$ to the best-fit value obtained for each device at the lowest temperature measured, we extract a value of $n_p$ at each measured temperature (Fig. \ref{percolation}). Using the functional form $n_p=n_0+C e^{-b/T}$, we can extrapolate the percolation threshold to zero temperature and extract $n_0$, the T=0 percolation threshold density.\cite{tracy} The exponential term $b$ is an energy gap related to the impurity distribution of the system. Our fit yields $n_p=0.83+1.46 e^{-2.25/T}$ for sample A and $n_p=0.95+0.88 e^{-3.00/T}$ for sample B, showing very similar T=0 percolation thresholds of $0.83\times 10^{11}$ cm$^{-2}$ and $0.95\times 10^{11}$ cm$^{-2}$, respectively.

\begin{figure}
\includegraphics[width=\linewidth]{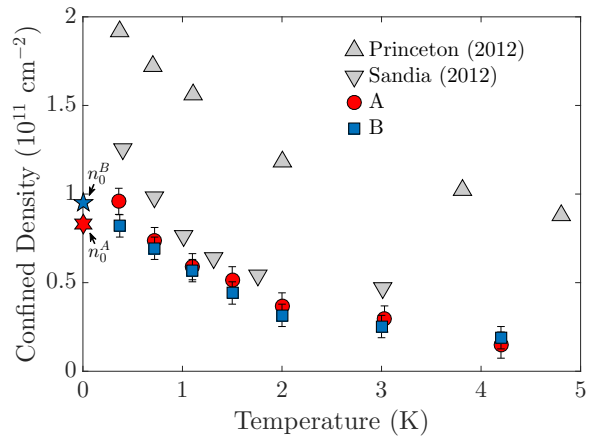}
\caption{\label{nconf} ESR measurement of density of electrons confined in shallow traps as a function of temperature for samples A and B compared to previously studied devices (triangles). T=0 percolation threshold densities for A (hexagram) and B (pentagram) are also plotted on the axis. Data from the previously studied devices (``Princeton (2012)" and ``Sandia (2012)") were reproduced from Appl. Phys. Lett. $\bf{100}$, 023503 (2012), with the permission of AIP Publishing.}
\end{figure}

Using ESR we can directly measured the density of electrons confined in shallow traps in samples A and B.\cite{jock,shankar} We use X-band ($\sim 9.6$ GHz, $\sim3400$ G) continuous wave ESR to measure the intensity of the 2DEG spin signal as a function of gate voltage ($V_G$) at fixed temperature between 360 mK and 4.2 K. Figure \ref{esr} shows an example of the number of unpaired spins, calculated as the double integral of the ESR spectrum, as a function of $V_G$. The data shown in Figure \ref{esr} is from sample B, measured at 1.50 K.  As $V_G$ is scanned below threshold, the ESR signal decreases as shallow traps in the channel are thermally depopulated. At some $V_G$, the signal saturates (``dark" curve in Fig. \ref{esr}) when the chemical potential is aligned with shallow traps deep enough that the confined electrons cannot thermally escape. We denote this characteristic voltage as $V^*$. Illuminating the sample with above band gap (1050 nm) light relaxes the system by neutralizing confined electrons with holes and the corresponding (``post-LED") ESR signal decreases and eventually goes to zero at voltage $V^0$.

With values for $V^*$ and $V^0$, we may then calculate the number of electrons confined in shallow traps ($n_{conf}$) at each measured temperature using the relation $e \cdot n_{conf} = C_{ox}(V^*-V^0)$, where $C_{ox}$ is the oxide capacitance measured from the Hall resistivity. The energy scale of the shallowest populated traps at each temperature\cite{jock,shankarprb} is approximately 1$0 k_B \cdot T$. Figure \ref{nconf} summarizes these data for samples A and B and plots data from a previous study for reference.\cite{jock} We note that the samples studied in the current work demonstrate similar densities of shallow traps to the previously studied Sandia device.\cite{jock} As temperature decreases, the density of confined charge increases as more electrons are frozen into shallower traps. Within experimental error, samples A and B demonstrate the same density of shallow traps across the measured temperature range, $\approx 9\times10^{10}$ cm$^{-2}$ at $\geq 0.3$ meV below $E_C$ (360 mK) and $\approx 3\times 10^{10}$ cm$^{-2}$ at $\geq 2$ meV below $E_C$ (2.0 K). This measurement is consistent with the T=0 percolation threshold densities for both devices studied ($n_0^{A,B}$), also plotted in figure \ref{nconf}, and demonstrates the efficacy of the forming gas anneal for removing e-beam generated shallow defects across the measured energy range.

In summary, we have fabricated and measured high-mobility MOSFETs and have shown that a forming gas anneal is sufficient to restore an e-beam irradiated sample to the quality of an un-irradiated sample, as measured by the the extraction of a T=0 percolation threshold and ESR measurements of the density of shallow traps. We believe that these measurements, as opposed to peak mobility, are more relevant metrics to characterize MOS interfaces for quantum devices operating in the low electron density regime and show that the two separate measurements agree with each other at the lowest measured temperature.

This work was supported by the NSF through the MRSEC Program (Grant No. DMR-01420541), and the ARO (Grant No. W911NF-13-1-0179). J.-S.K. is supported in part by the Program in Plasma Science and Technology at Princeton University.


%
%

%


\bibliography{references}

\end{document}